\documentclass[letterpaper, 10 pt, conference]{ieeeconf}  % Comment this line out if you need a4paper

\IEEEoverridecommandlockouts                              % This command is only needed if 
\usepackage{graphicx}
\usepackage{xcolor}
\definecolor{MyLightGreen}{rgb}{0.3, 0.6, 0.3}

\usepackage{multirow}
\usepackage{url}
\usepackage{graphicx}
\usepackage{graphics} % for pdf, bitmapped graphics files
\usepackage{amsmath,amssymb,amsfonts}

\usepackage{booktabs}
\usepackage{multirow}

\usepackage{pifont}
\usepackage{array}
\newcolumntype{L}[1]{>{\raggedright\arraybackslash}p{#1}}  % Define L{}

\newcommand{\cmark}{\textcolor{green!70!black}{\ding{51}}} % ✓
\newcommand{\xmark}{\textcolor{red}{\ding{55}}}            % ✗

\title{\LARGE \bf
CDA-SimBoost: A Unified Framework Bridging Real Data and Simulation for Infrastructure-Based CDA Systems
}

\author{Zhaoliang Zheng$^{\star}$$^{1}$, Xu Han$^{\star}$$^{1}$, Yuxin Bao$^{1}$, Yun Zhang$^{1}$, Johnson Liu$^{1}$, \\ Zonglin Meng$^{1}$, Xin Xia$^{2}$ and Jiaqi Ma$^{\dagger}$$^{1}$% <-this % stops a space
\thanks{$^{\star}$ Co-first Authors.}
\thanks{$^{1}$Zhaoliang Zheng, Xu Han, Yuxin Bao, Yun Zhang, Johnson Liu, Zonglin Meng, Jiaqi Ma are with
        University of California, Los Angeles, USA}%
\thanks{$^{2}$Xin Xia is with Department of Mechanical Engineering, University of Michigan-Dearborn, USA}%
\thanks{$^{\dagger}$ Corresponding Author: {\tt\small jiaqima@ucla.edu}}
}

\begin{document}

\maketitle
\thispagestyle{empty}
\pagestyle{empty}

%%%%%%%%%%%%%%%%%%%%%%%%%%%%%%%%%%%%%%%%%%%%%%%%%%%%%%%%%%%%%%%%%%%%%%%%%%%%%%%%
\begin{abstract}

Cooperative Driving Automation (CDA) has garnered increasing research attention, yet the role of intelligent infrastructure remains insufficiently explored. Existing solutions offer limited support for addressing long-tail challenges, real-synthetic data fusion, and heterogeneous sensor management. This paper introduces CDA-SimBoost, a unified framework that constructs infrastructure-centric simulation environments from real-world data. CDA-SimBoost consists of three main components: a Digital Twin Builder for generating high-fidelity simulator assets based on sensor and HD map data, OFDataPip for processing both online and offline data streams, and OpenCDA-InfraX, a high-fidelity platform for infrastructure-focused simulation. The system supports realistic scenario construction, rare event synthesis, and scalable evaluation for CDA research. With its modular architecture and standardized benchmarking capabilities, CDA-SimBoost bridges real-world dynamics and virtual environments, facilitating reproducible and extensible infrastructure-driven CDA studies. All resources are publicly available at \url{https://github.com/zhz03/CDA-SimBoost}.

\end{abstract}

%%%%%%%%%%%%%%%%%%%%%%%%%%%%%%%%%%%%%%%%%%%%%%%%%%%%%%%%%%%%%%%%%%%%%%%%%%%%%%%%
\section{Introduction}
In recent years, Cooperative Driving Automation (CDA) has gained significant traction \cite{parekh2022review}, leading to the development of a wide range of algorithms and system architectures to support its implementation. However, the role of intelligent infrastructure within CDA remains underexplored \cite{yu2022dair, zheng2025inspe, petrillo2024editorial}. Infrastructure-based sensing systems, due to their strategic placement and superior computational resources, offer unique advantages in CDA tasks—including perception, prediction, and decision-making. By providing elevated and extended viewpoints, these systems enhance the detection of occluded objects and enable broader situational awareness.

% Despite the emergence of several infrastructure-related datasets and studies \cite{yu2022dair, hao2024rcooper, xiang2024v2x}, most of these studies are aimed at data collection under routine circumstances and still face many key challenges. These include addressing the long-tail distribution of intersection scenarios, generating rare or complex events, fusing real and simulated data, managing heterogeneous sensor deployment, collaborative driving tasks, and deployment in challenging scenarios \cite{zheng2025inspe, scott2023scenario, su2025adopting}. 
% Digital twin technology provides a promising solution by enabling high-fidelity virtual intersections with flexible sensor and agent configurations, supporting safe, repeatable, and scalable testing of cooperative driving strategies.
Although several infrastructure datasets and studies have emerged \cite{yu2022dair, hao2024rcooper, xiang2024v2x}, most focus on routine data and still face key challenges. These include rare event generation, real-sim data fusion, heterogeneous sensors, cooperative driving, and deployment in complex scenarios \cite{zheng2025inspe, scott2023scenario, su2025adopting}. In addition, in real-world scenarios, intelligent infrastructures and autonomous vehicles will frequently operate in highly dynamic and safety-critical environments, such as occluded intersections, construction zones, emergency vehicle interactions, or near-miss conditions. However, these challenging edge cases are severely underrepresented in existing datasets and simulations, limiting the generalization and robustness of current CDA solutions. Moreover, collecting such data in the real world is expensive, dangerous, and logistically difficult. Digital twin technology provides a promising solution by enabling high-fidelity virtual intersections with flexible sensor and agent configurations, supporting safe, repeatable, and scalable testing of cooperative driving strategies under a wide range of conditions. 

%Digital twin technology offers a promising pathway to address these challenges by enabling the construction of high-fidelity virtual environments that faithfully replicate real-world intersections. Such environments allow flexible configuration of sensors and agents, and facilitate safe, repeatable, and scalable testing of cooperative driving strategies. By bridging the gap between real-world data and simulation, digital twins make it possible to integrate dynamic traffic flows, real-time perception, and controllable scenario generation within a unified simulation platform.

% To address these limitations and reduce the cost of infrastructure-centric CDA research, we present CDA-SimBoost, a unified framework for building digital twins based on real-world data. Building upon our prior work, OpenCDA \cite{xu2021opencda}, CDA-SimBoost supports both online and offline data ingestion and facilitates the creation of diverse, high-fidelity synthetic scenarios. The pipeline begins with constructing digital twins of real-world intersections using sensor data and High Definition (HD) maps. It features an online module for real-time streaming and an offline module for fine-grained processing. Both feed into Opencda-InfraX, our infrastructure-based simulation platform, which generates background and connected automated vehicle (CAV) data. Based on user-defined configurations, the platform synthesizes challenging multi-agent traffic scenarios to support scalable evaluation and training for intelligent traffic systems and CDA systems. 
To overcome these limitations and lower the cost of infrastructure-centric CDA research, we propose \textit{CDA-SimBoost}, a unified framework for constructing digital twins from real-world data. Extending OpenCDA \cite{xu2021opencda}, CDA-SimBoost supports both online and offline data ingestion to generate diverse, high-fidelity synthetic scenarios. It builds digital twins using sensor data and HD maps, with online modules for real-time input and offline modules for detailed processing. These feed into \textit{Opencda-InfraX}, our simulation platform for infrastructure-based CAV data generation. The platform synthesizes complex multi-agent scenarios for scalable evaluation and training of intelligent traffic and CDA systems. The main features of CDA-SimBoost can be summarized as follows: \textbf{Digital Twin Modeling:} Constructs high-fidelity digital twins from real-world data. \textbf{Data Import Modes:} Online and offline data processing pipelines support real-time ingestion and advanced offline analysis. \textbf{Modularity:} Enables independent use and extension of CDA components. \textbf{Full-Stack CDA System:} Provides infrastructure-centric CDA tools, including environment setup, sensors, and core modules (e.g., localization, perception, planning, communication). \textbf{Data Generation:} It integrates with flexible interfaces to generate synthetic and simulation data in customizable modes. \textbf{Benchmarking:} Includes benchmark scenarios that are based on the NHTSA report \cite{national2007pre}, and provides performance metrics on system and agent levels.

\begin{figure*}[!t]
    \centering
    \includegraphics[width=0.85\textwidth]{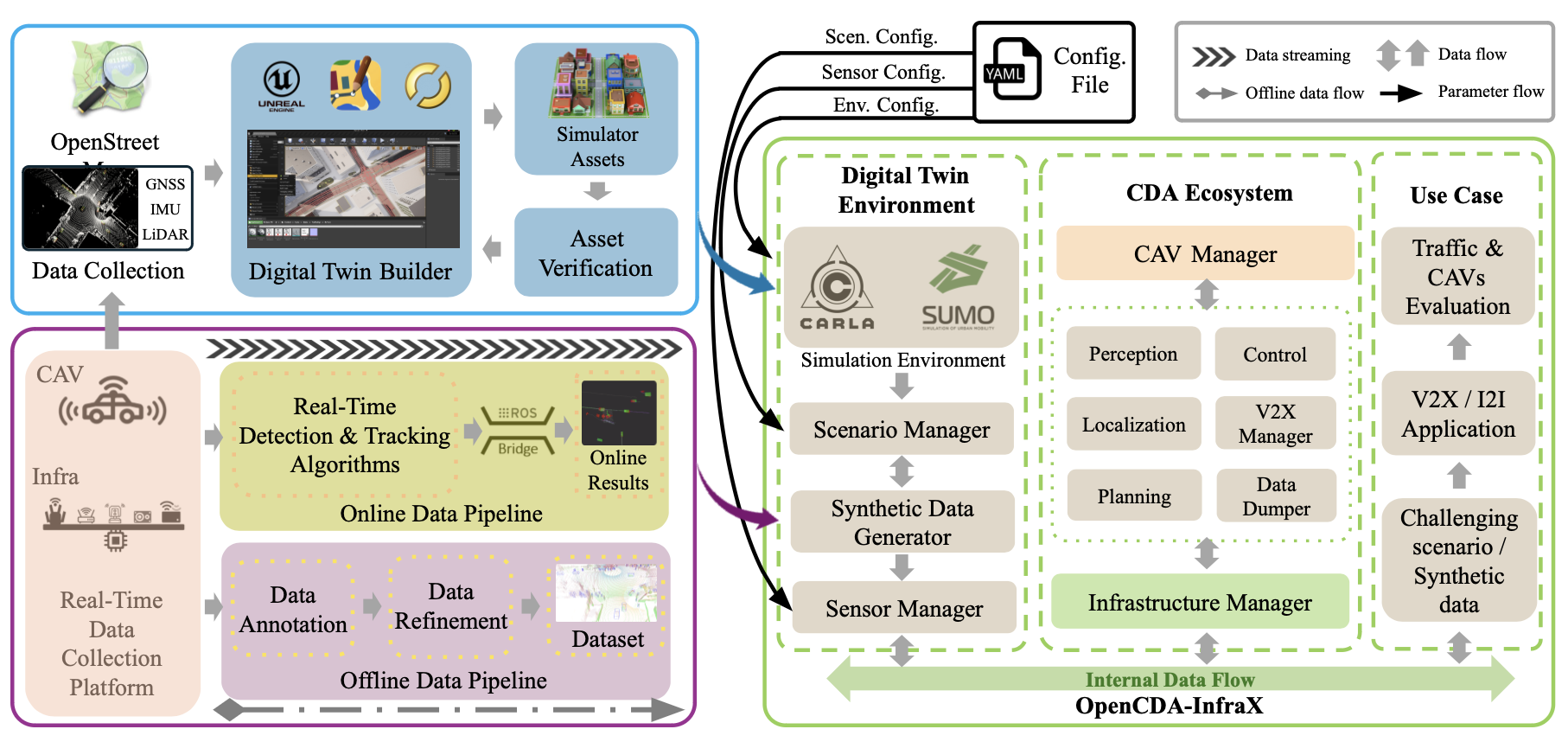}
    \caption{\textbf{Workflow of CDA-SimBoost framework.} There are three main parts in CDA-SimBoost: \textcolor{blue}{Digital Twin Builder}, \textcolor{purple}{Online-Offline Data Pipeline} and \textcolor{MyLightGreen}{OpenCDA-InfraX}. Digital Twin Builder provides simulator assets as input for the simulation platform. OFDataPip takes in real-world data collected by our data collection platform: connected automated vehicle (CAV) and smart infrastructure (Infra), and outputs converted data to OpenCDA-InfraX. The inputs of OpenCDA-InfraX also include configuration file (Config. File) that composes scenario configuration (Scen. Config.), sensor configuration (Sensor Config.), as well as environmental configuration (Env. Config.). V2X, I2I stands for vehicle-to-everything and infrastructure-to-infrastructure, respectively.}
    \label{fig:1}
    \vspace{-5mm}
\end{figure*}

% \begin{itemize}
%     \item Digital Twin Modeling: Constructs high-fidelity digital twins from real-world data.
    
%     \item Data Import Modes: Online and offline data processing pipelines support real-time ingestion and advanced offline analysis.
    
%     % \item Modularity: Enables independent use and extension of components like the twin builder, real-time detection, and OpenCDA-InfraX modules.
%     \item Modularity: Enables independent use and extension of CDA components.

%     \item Full-Stack CDA System: Provides infrastructure-centric CDA tools, including environment setup, sensors, and core modules (e.g., localization, perception, planning, communication).

%     \item Data Generation: It integrates with flexible interfaces to generate synthetic and simulation data in customizable modes.
    
%     \item Benchmarking: Includes benchmark scenarios that are based on the NHTSA report \cite{national2007pre}, and provides performance metrics on system and agent levels.
% \end{itemize}

\section{Related Work}
% \textcolor{red}{Yuxin Bao NEEDS REVISION}

% Several recent frameworks and simulators have been developed to advance autonomous vehicle research with a focus on data collection, simulation fidelity, and scenario generation. However, most of these tools are centered on single-vehicle behavior and offer limited support for cooperative driving automation (CDA). Scenario Gym \cite{scott2023scenario} is a lightweight, scenario-driven simulator that enables rapid prototyping of complex driving scenarios through customizable agents. GarchingSim \cite{zhou2023garchingsim}, built on the Unity Engine \cite{unity}, provides photorealistic 3D environments, accurate vehicle dynamics, and configurable sensor setups. Despite their strengths, both simulators lack support for cooperative interactions and infrastructure-based scenario modeling, limiting their applicability to CDA-focused research. 

Several recent frameworks and simulators have been developed to advance autonomous vehicle research, with a particular emphasis on data collection, simulation fidelity, and scenario generation \cite{seiffer2024control, xiang2022v2xp}. However, most existing tools focus primarily on single-vehicle behavior and provide limited support for CDA. For example, Scenario Gym \cite{scott2023scenario} is a lightweight, scenario-driven simulator that facilitates rapid prototyping of complex driving scenarios using customizable agents. GarchingSim \cite{zhou2023garchingsim}, built on the Unity Engine \cite{unity}, offers photorealistic 3D environments, accurate vehicle dynamics, and configurable sensor setups. While both simulators offer notable strengths, they lack support for cooperative interactions and infrastructure-based scenario modeling, thereby limiting their applicability to CDA-focused research. In parallel, digital twin frameworks have emerged to bridge the physical and digital worlds for autonomous driving. Digital Twins for Autonomous Driving \cite{wang2024digital, kabir2025digital, martinez2025sustainable} presents an end-to-end real-world system combining roadside sensing, edge computing, and cloud-based route planning. Another digital twin platform, Astral 3D Editor \cite{astral}, provides web-based collaboration for general-purpose modeling. Nevertheless, Digital Twins for Autonomous Driving focus on route optimization rather than serving as a general platform for CDA research, while Astral lacks the domain-specific capabilities, such as real-time traffic integration and infrastructure-agent coordination. Complementing this direction, OpenCDA \cite{xu2021opencda} offers a full-stack simulation framework for CDA. Its extension, OpenCDA Open-Source Ecosystem \cite{xu2023opencda}, enhances the platform with multi-resolution simulators, a model zoo, and scenario databases. Despite these capabilities, the OpenCDA ecosystem is largely synthetic and lacks support for real-time integration of live sensor input, limiting its ability to model continuously evolving, real-world traffic environments. In light of these limitations, we propose CDA-SimBoost, a real-time digital twin framework. By enabling infrastructure-integrated real-time simulation, CDA-SimBoost bridges the gap between static simulations and real-world traffic dynamics.

To further highlight the advantages of CDA-SimBoost, we compare it against representative simulators and frameworks in Table~\ref{tab:cda_comparison}. Unlike existing tools, CDA-SimBoost uniquely integrates real-world data streaming, high-fidelity digital twin modeling, infrastructure-agent coordination, and a full-stack CDA ecosystem. This unified design enables reproducible, scalable, and infrastructure-driven research that bridges the gap between simulation and deployment.

\begin{table}[h]
\centering
\scriptsize
\caption{
Comparison of simulation frameworks on CDA-relevant capabilities.}
\renewcommand{\arraystretch}{1.2}
\resizebox{\columnwidth}{!}{%
\begin{tabular}{|p{4.1cm}|c|c|c|c|}
\hline
\textbf{Methods} & \textbf{GarchingSim\cite{zhou2023garchingsim}} & \textbf{Scenario Gym\cite{scott2023scenario}} & \textbf{OpenCDA\cite{xu2023opencda}} & \textbf{Ours} \\
\hline
Infrastructure-agent Modeling \& Sensing Support & \xmark & \xmark & \xmark & \cmark \\
\hline
ROS Integration & \cmark & \xmark & \xmark & \cmark \\
\hline
Digital Twin Capability & \xmark & \xmark & \cmark & \cmark \\
\hline
Real-time Data Support & \xmark & \xmark & \xmark & \cmark \\
\hline
Scenario Flexibility \& Customization & \xmark & \cmark & \cmark & \cmark \\
\hline
Number of Multi-Class Types (Pedestrians, Cyclists, etc.) & 2 & 2 & 5+ & 5+ \\
\hline
Number of Communication Modes (V2X/V2V/V2I/I2I) & 2 & 0 & 3 & 4 \\
\hline
\end{tabular}
}
\label{tab:cda_comparison}
\vspace{-5mm}
\end{table}

% requirements: \begin{itemize}
%     \item 0.5 - 0.75 page
%     \item find out the most relevant papers and discuss their work and limitations. 1 sentence content (good) + 1 sentence limitation. 
%     \item multiple papers are solving the same topic, combine them and talk about their work.
%     \item time order or evolution order, 
%     \item after you finish talking about their limitations, start to mention our work. 1-2 sentences. 
% \end{itemize}

% some reference paper:
% \begin{itemize}
%     \item OpenCDA: An Open Cooperative Driving Automation Framework  Integrated with Co-Simulation
%     \item The OpenCDA Open-source Ecosystem for  Cooperative Driving Automation Research
%     \item How Simulation Helps Autonomous Driving:  A Survey of Sim2real, Digital Twins, and Parallel Intelligence
%     \item Scenario Gym: a scenario-centric lightweight simulator
%     \item GarchingSim: An Autonomous Driving Simulator with Photorealistic  Scenes and Minimalist Workflow
%     \item Digital Twins for Autonomous Driving: A  Comprehensive Implementation and Demonstration
%     \item Waymax: An accelerated simulator for autonomous driving research. (This is a Github, go search and check the reference.)
% \end{itemize}

\section{Overview of CDA-SimBoost}
\begin{figure*}[!t]
    \centering
    \includegraphics[width=0.9\textwidth]{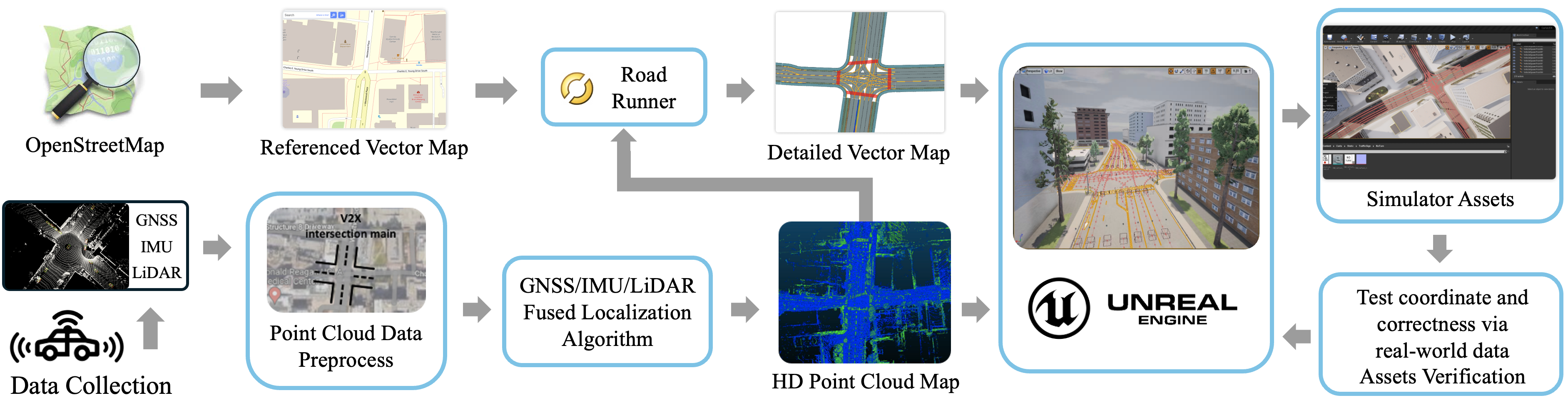}
    \caption{Workflow of Digital Twin Builder.}
    \vspace{-5mm}
    \label{fig:DTB}
    \vspace{-2mm}
\end{figure*}

CDA-SimBoost is a general-purpose framework that integrates a digital twin builder, online and offline data ingestion pipelines (OFDataPip), and an infrastructure-centric cooperative driving simulation platform—OpenCDA-InfraX. 
% Entirely developed in Python, it supports digital twin construction, real-world data integration, scenario generation, and the development and validation of CDA algorithms. 
This section introduces the digital twin builder and OFDataPip modules, which are key to preparing and ingesting real-world data into OpenCDA-InfraX. The overall workflow is illustrated in Fig. \ref{fig:1}.

\subsection{Digital Twin Builder (DTB)}
The digital twin of the targeted intersection serves as a virtual replica of a real-world site, designed for deployment within the simulation environment \cite{dosovitskiy2017carla}. As shown in Fig. \ref{fig:1}, the goal of DTB is to generate digital assets for OpenCDA-InfraX. To construct the simulation, digital assets derived from real-world data are imported into Unreal Engine \cite{engine2018unreal}, where the final CARLA map is generated. Two types of digital assets are essential: a HD point cloud map that captures detailed 3D spatial information, and a vector map \cite{opendrive} that defines roadway geometry, topology, and lane configurations.  

To generate the HD point cloud map, a data collection vehicle equipped with a 128-line LiDAR, an inertial measurement unit (IMU) and a Global Navigation Satellite System (GNSS) sensor suite is deployed to scan the targeted intersection. The vehicle traverses the area, capturing multiple LiDAR frames synchronized with GNSS data enhanced by correction methods \cite{dabove2019towards}. Firstly, the raw LiDAR points are transformed into UTM coordinates based on the vehicle’s GNSS-reported position and orientation. Points from irrelevant regions (e.g., vegetation or road surface), identified with reference to satellite imagery, are flagged and removed to reduce data size. We leverage a GNSS/IMU/LiDAR fusion-based localization algorithm \cite{gao2023gnss} to aggregate all processed frames, resulting in a compact and HD point cloud map (Fig. \ref{fig:DTB}) suitable for efficient simulation. 

After finalizing the HD point cloud map, a high-definition vector map is manually generated by referencing a simplified OpenStreetMap vector map and tracing roadway and lane geometries, along with detailed annotations, over the 3D point cloud using MathWorks RoadRunner \cite{roadrunner}. RoadRunner ensures proper connectivity between road segments before exporting the vector map for use in Unreal Engine. CARLA then processes the vector map to generate waypoints and define drivable areas. Simulation assets such as road surfaces, materials, vegetation, and objects are assigned according to CARLA’s standards and real-world references. Environmental elements, including weather, lighting, and traffic signals, are also configured to reflect realistic intersection behavior. The resulting CARLA assets and workflow are shown in Fig. \ref{fig:DTB}. The process concludes with multiple validation runs to ensure full functionality before exporting a standalone, distributable map package.  
\begin{figure}[h]
    \centering
    \includegraphics[width=0.7\linewidth]{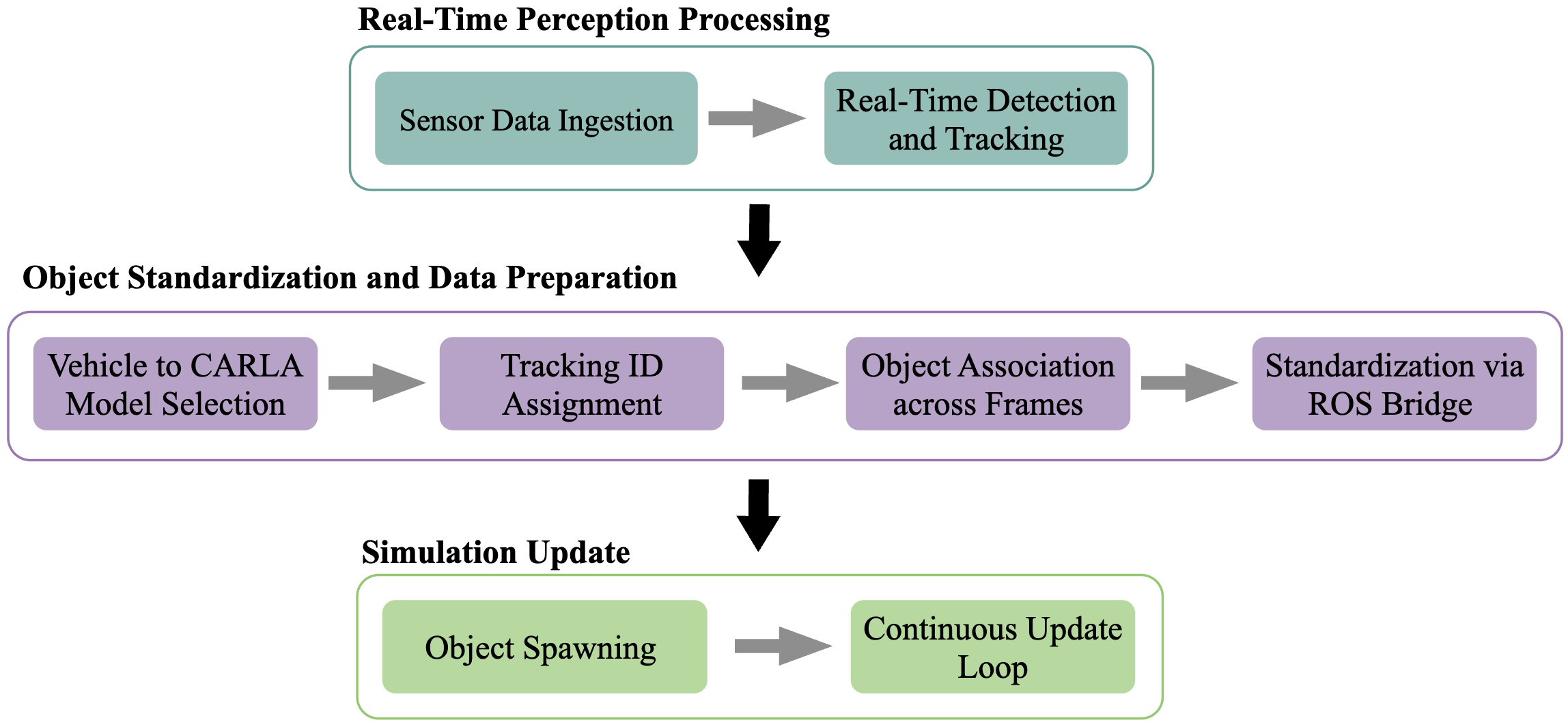}
    \caption{A flowchart of the online data processing pipeline in OFDataPip.}
    \vspace{-3mm}
    \label{fig:onlinedata}
    \vspace{-3mm}
\end{figure}

\subsection{Online-Offline Data Pipeline (OFDataPip)}

% In this section, we will introduce how to process real-time data, first real-time detection and tracking algorithms, then via ROS bridge, convert data into CARLA. 

As shown in Fig. \ref{fig:1}, the purpose of OFDataPip is to collect and generate real-time data streams or high-quality offline data for OpenCDA-InfraX. For the offline data pipeline, we leverage data labeling technique and refinement to get better quality data, including more diverse labels, more accurate bounding boxes, and improved consistency in tracking information across frames. The offline data annotation module adopts a hybrid strategy combining model-assisted pre-labeling with manual annotation, achieving high labeling accuracy while improving efficiency. To further enhance annotation quality, we developed a refinement algorithm based on trajectory consistency, which adjusts the annotation results to ensure coherent object trajectories and consistent tracking IDs across frames. To support real-time data processing, we developed an online pipeline that integrates raw data streams from infrastructure-mounted devices, connected autonomous vehicles (CAVs), and other real-time sources. As shown in Fig. \ref{fig:onlinedata}, incoming data is processed by object detection and tracking algorithms \cite{xia2023automated} that extract attributes such as position, orientation, and bounding dimensions for each detected object. To maintain consistency within digital twin, vehicle models are selected based on the closest size match from the simulation asset library. Each detected object is assigned a unique tracking ID, and multi-frame association algorithms ensure consistent labeling across frames, enabling accurate trajectory reconstruction. The module is highly modular, with detection algorithms and data sources decoupled from the publishing process to the digital twin environment. After detection, object states—including position, orientation, type, and tracking ID—are standardized into a unified message format using the ROS bridge package, enabling seamless integration with the digital twin simulation platform. As new object states arrive in subsequent frames, updates to existing actors (e.g., position and orientation) are sent via the ROS bridge \cite{zheng2023opencda}. This online pipeline allows CDA-SimBoost to maintain a real-world data stream within OpenCDA-InfraX.

\section{OpenCDA-InfraX}
\begin{figure}[!h]
    \centering
    \includegraphics[width=0.49\textwidth]{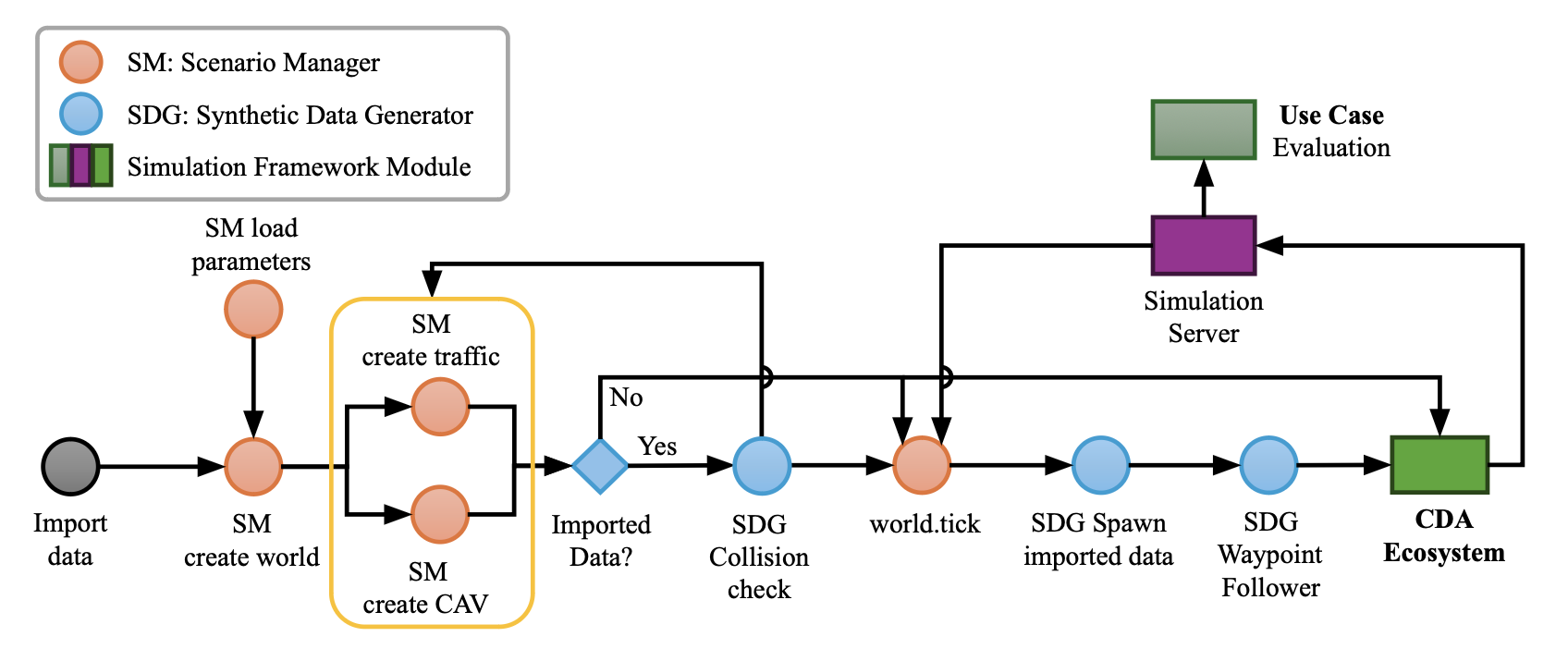}
    \caption{\textbf{The logic flow of OpenCDA-InfraX.} The circular elements in different colors represent their affiliation with different subsystem modules, while the square elements denote composite subsystems.
}
    \vspace{-3mm}
    \label{fig:logic_flow1}
    \vspace{-3mm}
\end{figure}

Built upon the design framework of OpenCDA \cite{xu2021opencda}\cite{xu2023opencda}, \textit{OpenCDA-InfraX} is an infrastructure-centric cooperative driving simulation platform that integrates multiple components, including a digital twin environment, a comprehensive CDA ecosystem, and use cases. The detailed logic flow of the simulation process is illustrated in Fig. \ref{fig:logic_flow1}.

\subsection{Digital Twin Environment}

In \textit{OpenCDA-InfraX}, the digital twin environment is essential for configuring simulation settings. Users initialize the simulator by importing assets from the DTB—including roads, buildings, and infrastructure—which form the simulation’s foundation. Configuration files allow customization of parameters such as weather, lighting, time units, and simulation duration. The \textit{scenario manager} in \textit{OpenCDA-InfraX} is responsible for generating dynamic simulation scenarios that describe how the virtual environment evolves. Users define scenario configurations that specify traffic flows, background road users generation, and other dynamic behaviors. Notably, these configurations can function as complete, standalone simulation scenarios. Currently, \textit{OpenCDA-InfraX} includes approximately ten predefined testing scenarios, each executable in CARLA and capable of producing corresponding datasets. The dynamic elements generated by the scenario manager are designed to interact seamlessly with real-world data sources. To facilitate this interaction, we introduce a \textit{Synthetic Data Generator}, as illustrated in Fig. \ref{fig:logic_flow1}. This interface ingests real-world inputs and generates interactions between real-world data and simulated scenario elements, with a particular focus on detecting potential conflicts and collisions. Through this mechanism, the system synthesizes hybrid datasets that integrate both real and simulated data, supporting the generation of high-fidelity, diverse scenarios. The resulting datasets encompass multiple object classes, including vehicles, pedestrians, trucks, buses, and cyclists, thereby enriching the representativeness and applicability of the synthetic data for cooperative driving research.  

% \textcolor{blue}{talk about the waypoint tracker.}

% \textcolor{blue}{In the scenario manager, talk about the traffic generation: pedestrians, truck, bus, cyclist.}

% Here I need to have one figure to show the logic flow of synthetic data generator. 

\subsection{CDA Ecosystem}

In contrast to OpenCDA \cite{xu2021opencda}, which primarily centers on autonomous vehicles, \textit{OpenCDA-InfraX} adopts an infrastructure-centric architecture by modeling infrastructure components as intelligent agents endowed with cooperative driving capabilities. 
% Within the CDA ecosystem of \textit{OpenCDA-InfraX}, the infrastructure-centric definition and subsystem design constitute the primary distinctions from OpenCDA \cite{xu2021opencda}.
% This paradigm leverages the intrinsic advantages of infrastructure in intelligent transportation systems, such as powerful sensing, and strategically elevated placements, which allow infrastructure sensors to ``stand higher, see farther, and perceive more clearly". 

\noindent \textbf{Infrastucture Definition.} Infrastructure Manager is introduced to implement this infrastructure-centric paradigm, enabling each infrastructure unit to be defined, managed, and operated as an autonomous agent. Components co-located at the same intersection are further organized into an intersection-level agent group, supporting coordinated perception and decision-making across the local infrastructure system. We define intelligent intersections region $\mathcal{I}$ in details as follows:

\begin{equation}
    \mathcal{I} = \left\{ 
    % (x, y, z) \in \mathbb{R}^3 \;\middle|\; 
    \begin{aligned}
    \sqrt{(x - x_c)^2 + (y - y_c)^2} \leq d_f, \\ \text{ground} \leq z \leq \text{ground} + 4 \,\text{m}
    \end{aligned}
    \right\},
\end{equation}
where, $x_c,y_c,z_c$ is the center of the intersection and $(x, y, z) \in \mathbb{R}^3$ is the location point within the intersection region. $d_f$ is the radius of the region and typically ranges from 50 to 100 meters. 

The individual infrastructure unit (IU) is defined as:
\begin{equation}
\text{IU} = \left\{ s \in \mathcal{S} \ \middle|\ 
\begin{aligned}
&\forall\, s_i, s_j \in \text{IU}, \\
&\sqrt{(x_i - x_j)^2 + (y_i - y_j)^2} \leq 2m, \\
&|z_i - z_j| \leq 4m, \\
& p_i = p_j
\end{aligned}
\right\} , 
\end{equation}
where, $s$ represents a sensor, $i,j$ are different sensor indices, $\mathcal{S}$ is the set of all sensors in one IU, and $x,y,z$ represent the physical location of a sensor. The $p_i = p_j$ ensures that all sensors in the IU share the same processing unit.  

\noindent \textbf{Subsystem Modules.} \textit{Infrastructure Manager} provides a modular framework for defining and managing infrastructure-mounted sensors through the integrated \textit{Sensor Manager}, enabling flexible configuration based on application needs. Shared functional modules such as \textit{Perception}, \textit{V2X Manager}, and \textit{Planning} are reused across both the \textit{CAV Manager} and \textit{Infrastructure Manager}, while components like \textit{Control}, \textit{Localization}, and \textit{Data Dumper} are adapted to reflect the distinct characteristics of infrastructure units (IUs). For example, the \textit{Control} module supports infrastructure-specific actions such as traffic light phase management, the \textit{Localization} module leverages GPS for IU positioning, and the \textit{Data Dumper} allows for customized data output per IU.

\noindent \textbf{Modular Libraries.} OpenCDA-InfraX offers a comprehensive suite of cooperative driving applications, encompassing both rule-based and learning-based approaches. These applications are fully modular, enabling users to easily replace, extend, or customize individual components to support diverse development and research requirements. Currently, the modular libraries include: \textit{LiDAR Object Detection:} VoxelNet \cite{Zhou_2018_CVPR}, PointPillar \cite{Lang_2019_CVPR}, SECOND \cite{yan2018second}. \textit{Camera Object Detection:} YOLO \cite{Redmon_2016_CVPR}, Lift-Splat-Shot \cite{philion2020lift}. \textit{Cooperative Perception:} V2VNet \cite{wang2020v2vnet}, AttFuse \cite{xu2022opv2v}, V2X-ViT \cite{xu2022v2x}, where2Comm \cite{hu2022where2comm}, DiscoNet \cite{Mehr_2019_ICCV}. \textit{Object Tracking:} Kalman Filter + Hungarian Matching \cite{kuhn1955hungarian}. \textit{Cooperative Fusion:} late\cite{zheng2024cooperfuse}, intermediate \cite{xu2022opv2v}. \textit{Localization:} Kalman Filter Series \cite{welch1995introduction,chui2017kalman, wu2022joint}, GNSS/IMU/LiDAR Fuse \cite{gao2023gnss}. \textit{Prediction:} V2X-PNP \cite{zhou2024v2xpnp}. \textit{Planning:} A* searching \cite{hart1968formal}, Cubic Spline \cite{de1978practical, zheng2020point}. \textit{Control:} PID control \cite{ang2005pid}, MPC \cite{garcia1989model}.

% \begin{itemize}
%     \item LiDAR Object Detection: VoxelNet \cite{Zhou_2018_CVPR}, PointPillar \cite{Lang_2019_CVPR}, SECOND \cite{yan2018second}.
%     \item Camera Object Detection: YOLO \cite{Redmon_2016_CVPR}, Lift-Splat-Shot \cite{philion2020lift}.
%     \item Cooperative Perception: V2VNet \cite{wang2020v2vnet}, AttFuse \cite{xu2022opv2v}, V2X-ViT \cite{xu2022v2x}, where2Comm \cite{hu2022where2comm}, DiscoNet \cite{Mehr_2019_ICCV}. 
%     \item Object Tracking: Kalman Filter + Hungarian Matching \cite{kuhn1955hungarian}. 
%     \item Trajectory Tracking: Pure Pursuit \cite{coulter1992implementation}, MPC \cite{garcia1989model}, PID \cite{ang2005pid}.  
%     \item Localization: Kalman Filter Series \cite{welch1995introduction,chui2017kalman, wu2022joint}, GNSS/IMU/LiDAR Fuse \cite{gao2023gnss}.
%     \item Prediction: V2X-PNP \cite{zhou2024v2xpnp}.
%     \item Planning: A* searching \cite{hart1968formal}, Cubic Spline \cite{de1978practical}. 
%     \item Control: PID control \cite{ang2005pid}, MPC \cite{garcia1989model}.
% \end{itemize}

% I also need to have one figure to show the logic flow of simulation process of OpenCDA-InfraX. 

% Talk about the model libraries. 

\subsection{Use Case Design and Evaluation}

\textit{OpenCDA-InfraX} supports synthetic data generation, V2X/I2I applications, sensor placement evaluation and evaluation of both traffic systems and individual CAVs. It includes targeted scenarios that model interactions between CAVs and infrastructure units, incorporating realistic communication models and delay simulations to reflect real-world transmission characteristics. The I2I application further enables data coordination across multiple intersections within a corridor, supporting complex multi-intersection cooperative behaviors. The framework enables users to test various sensor configurations—such as camera resolution, LiDAR range, and radar angles—and assess trade-offs between cost and performance, facilitating systematic comparison of different deployment methods. In the use case evaluations, the framework provides safety- and stability-oriented metrics for traffic-level analysis, along with detailed evaluation criteria for individual CAVs tailored to various inspection tasks and operational scales. 

In terms of traffic-level evaluation, several key metrics are considered: throughput, defined as the number of vehicles successfully traversing the intersection within a specified time interval, serves as a measure of traffic capacity; delay, characterized by the average and maximum waiting times experienced by vehicles at control points, reflects congestion levels and signal timing efficiency; and average speed, computed across all vehicles within the region of interest, indicates the smoothness and efficiency of traffic flow. Together, these metrics provide a comprehensive assessment of system-level performance under varying traffic conditions and control strategies. For individual CAV evaluation, we provide the following metrics: 1) Detection metrics based on average precision (AP) across different Intersection over Union (IoU) thresholds, and multiclass mean Average Precision (mAP). 2) A distance-based mAP metric following the nuScenes evaluation standards, specifically designed to accommodate small objects such as pedestrians and cyclists. 3) Pose estimation metrics, including Average Translation Error (meters), Average Scale Error ($1 - \text{scale IoU}$), and Average Orientation Error (radians), to assess perception performance under various conditions.

\begin{figure}[!h]
    \centering
    \includegraphics[width=0.45\textwidth]{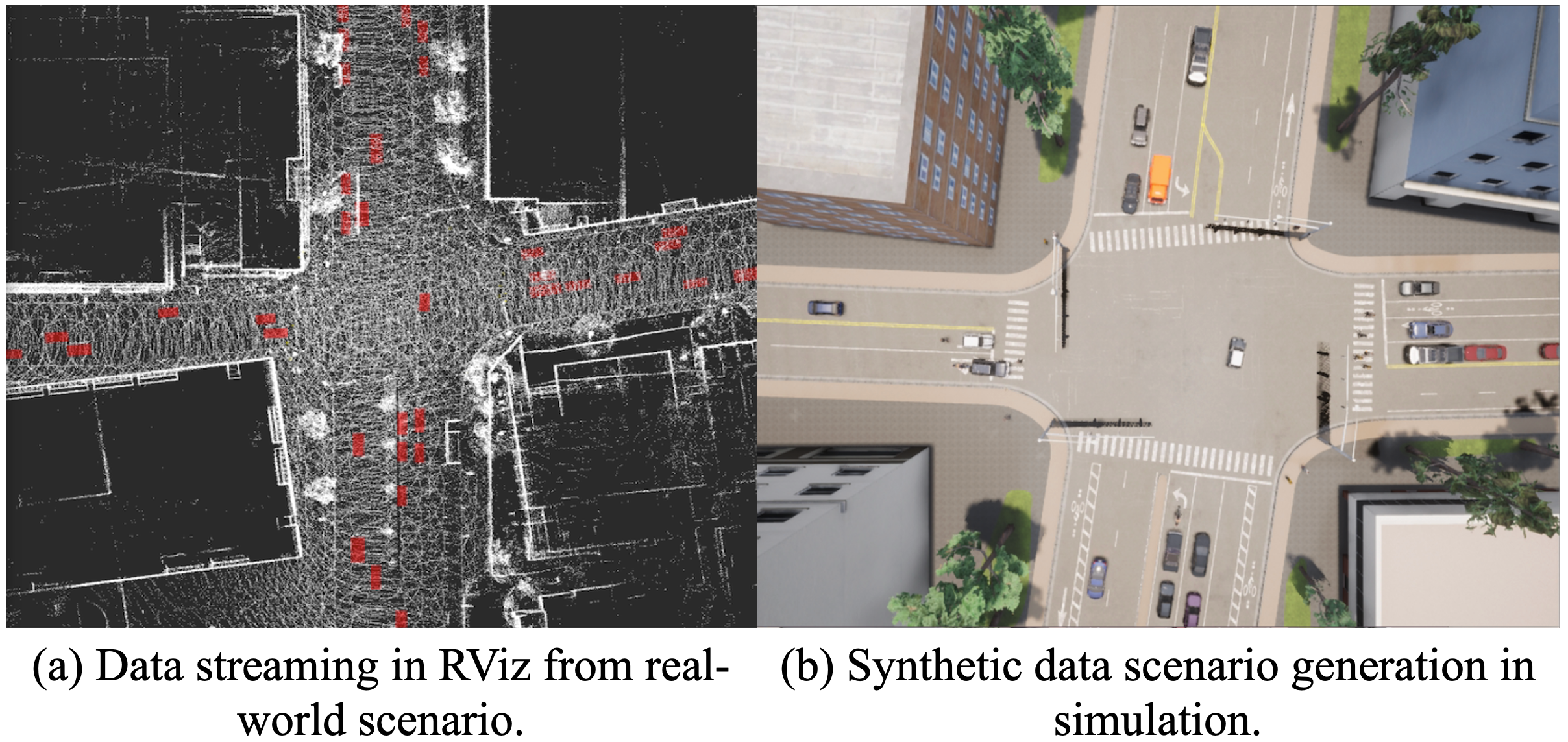}
    \caption{ \textbf{Digital Twin Data Streaming Visualization.} (a) shows real-world vehicle data detected by the real-time detection and tracking module, while (b) presents the synthetic data in the simulation environment based on the real-world data stream.
    }
    \vspace{-3mm}
    \label{fig:data_streaming}
    \vspace{-3mm}
\end{figure}

\section{Experiments}
% \begin{itemize}
%     \item challenging synthetic data/scenarios generation based on real-world data
%     \item sensor placement evaluation 
%     \item cav behavior level interaction (Xu) 
%     \item traffic level application (Xu) 
% \end{itemize}
\begin{figure}[!h]
    \centering
    \includegraphics[width=0.49\textwidth]{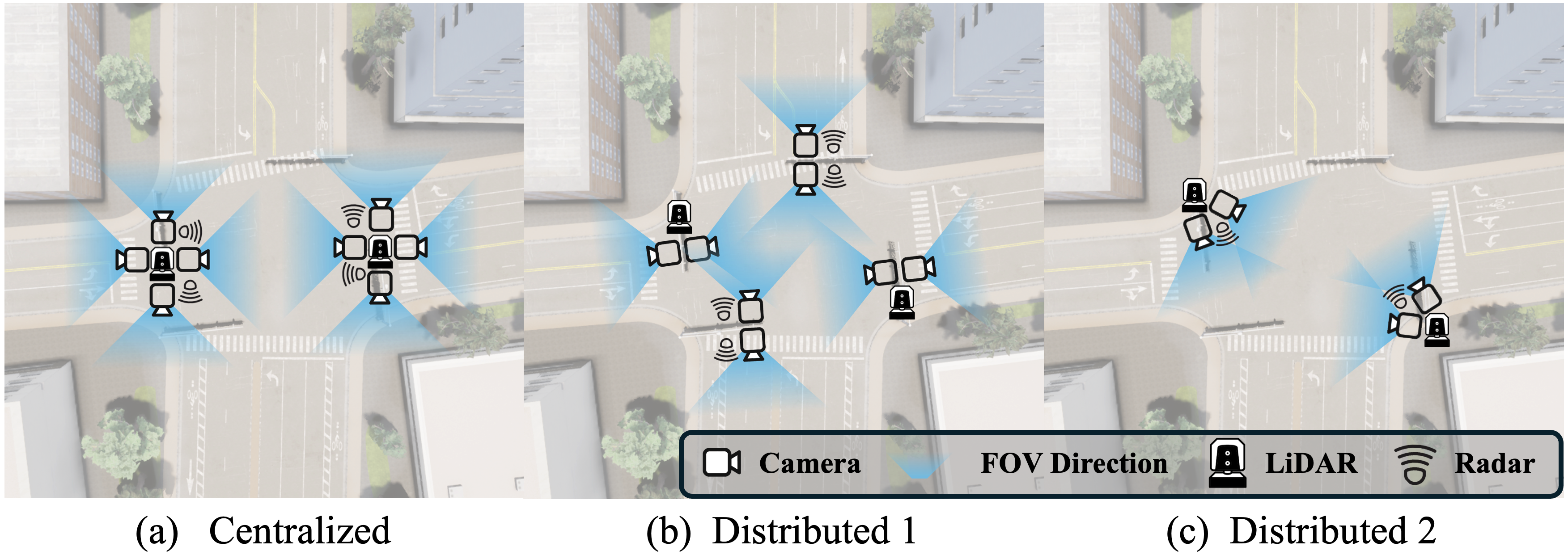}
    \caption{\textbf{Illustration figure of three types of sensor placements.} (a) features sensors concentrated near the center of the intersection, whereas (b) and (c) employ a more dispersed placement throughout the intersection. The camera arrangement in (a) is similar to that of the V2XSet \cite{xu2022v2x} dataset, (b) resembles those in the DAIR-V2X \cite{yu2022dair} and RCooper \cite{hao2024rcooper} datasets, and (c) is akin to the V2X-Real  \cite{xiang2024v2x} dataset. FOV direction is the camera's field of view direction.}
    \vspace{-3mm}
    \label{fig:sensor_placement}
    \vspace{-3mm}
\end{figure}

\begin{figure*}[!h]
    \centering
    \includegraphics[width=0.99\textwidth]{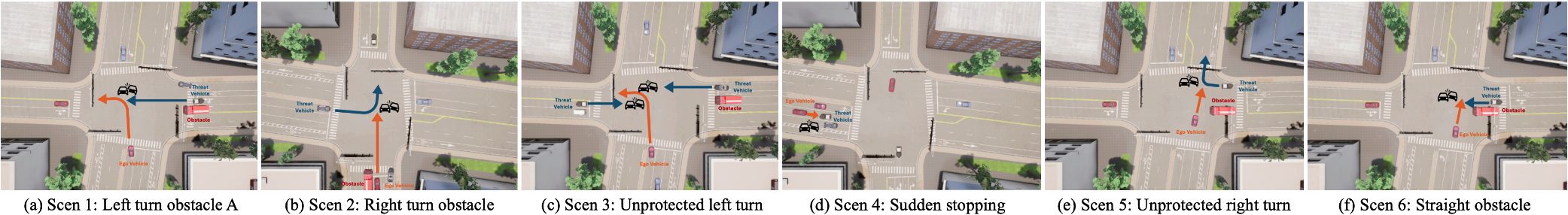}
    \caption{\textbf{Challenging synthetic scenario visualization.} (a)$\sim$(f) illustrate six different potential pre-crash scenario moments, each representing a simplified version generated by the \textit{Synthetic Data Generator} based on real-world data. In these scenarios, the ego vehicle is controlled by the CDA system of OpenCDA-InfraX.}
    \vspace{-3mm}
    \label{fig:challenging_scenario}
    \vspace{-3mm}
\end{figure*}

To evaluate the proposed framework, we conduct experiments targeting its support for real-world data integration, scenario customization, full-stack CDA functionality, and performance evaluation. Each experiment begins with real-world background traffic loading, followed by interactive synthetic scenario generation where an controllable ego vehicle navigates a predefined route. 
% Repeated simulation runs produce both system-level and agent-level performance metrics and visualizations, demonstrating the digital twin capability and framework’s effectiveness in end-to-end CDA evaluation.

\subsection{Challenging Synthetic Data Generation}

\noindent \textbf{Digital Twin Data Stream.}
In the synthetic data generation process, real-world sensor streams collected from the target location are first processed through the OFDataPip module and projected into the simulation environment to replicate actual traffic conditions. These data streams are imported via the ROS bridge, and support synchronized playback using ROSbag \cite{zheng2023opencda}. As illustrated in Fig. \ref{fig:data_streaming}, the left side shows real-world sensor observations, while the right displays their projection in simulation. Additional simulated objects and scenarios can be introduced on top of this foundation to enrich the environment and support diverse testing needs.

\noindent \textbf{Challenging Synthetic Scenario Visualization.}
Based on NHTSA analytical data and guidelines, and in combination with the real-world data streams, we have generated a variety of challenging scenarios that correspond to rare, long-tail events that are difficult to detect in real-world settings. Fig. \ref{fig:challenging_scenario} illustrates the collision moments in synthetic challenging scenarios, showcasing different dangerous driving situations. In CDA-SimBoost, we have designed six types of hazardous use cases, which users can customize more use cases according to their specific requirements. In addition, we provide examples of traffic violation scenarios implemented in the simulated town environments, enabling users to test their downstream algorithms under diverse and realistic conditions.

% It has two modes, pure simulation data generation and mixed real-world data + synthetic data generation. 

\subsection{Performance Analysis} 
 
% To facilitate the evaluation of sensor deployments and configurations, our CDA-SimBoost framework allows users to customize and configure various sensor parameters. As illustrated in Figure \ref{fig:sensor_placement}, our framework supports sensor placement and definition methods referenced from multiple datasets and can output dataset formats corresponding to whether sensors are colocated within the same infrastructure unit (IU). This design enables flexible evaluation across diverse sensor placement strategies.
\noindent \textbf{Sensor Placement Evaluation Capability.} To support detailed analysis of sensor deployment strategies, the CDA-SimBoost framework enables flexible sensor configuration within a digital twin environment. As shown in Fig. \ref{fig:sensor_placement}, the framework supports multiple sensor placement schemes, including centralized (a), partially distributed (b), and fully distributed (c) layouts. Each configuration can be populated with a customizable mix of cameras, LiDARs, and radars, along with their sensor types, positions, fields of view and orientation. This setup allows for systematic evaluation of perception coverage, occlusion handling, and redundancy under various deployment strategies, offering a practical tool for optimizing real-world infrastructure sensor layouts.

\begin{figure}[!h]
    \centering
    \includegraphics[width=0.40\textwidth]{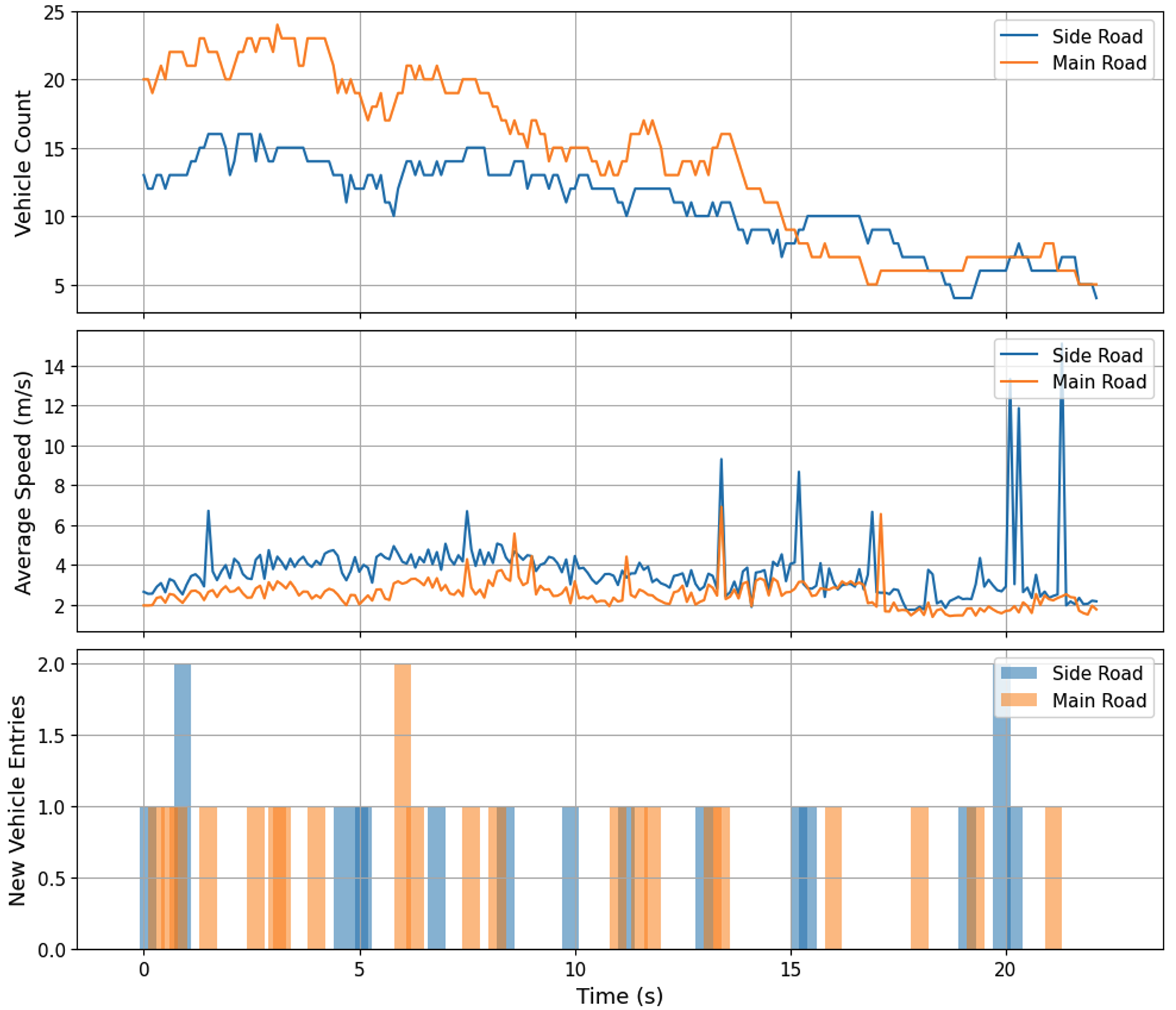}
    \vspace{-3mm}
    \caption{Traffic analysis comparison of main road and side road. This figure shows the vehicle count at the intersection (top), average speed of vehicles at the intersection (middle) and new vehicle entries to the intersection (bottom). 
    }
    \label{fig:traffic_analysis}
    \vspace{-4mm}
\end{figure}

\noindent \textbf{System-Level Performance Analysis.} 
The framework enables analysis of system-level traffic dynamics. leveraging infrastructure side sensors (i.e., Lidar, Cameras, and Radars). These edge-side sensors allow the proposed framework to capture and utilize metrics such as vehicle count, average speed, and vehicle entry rate, as illustrated in Figure \ref{fig:traffic_analysis}. The distinct trends observed between the main and side roads, such as the consistently higher vehicle count and throughput on the main road, highlight the framework’s ability to capture realistic traffic variations and congestion effects. This level of simulation fidelity is crucial for evaluating intersection performance, traffic coordination strategies, and policy interventions in a controlled environment. By replicating the flow and behavior patterns of the physical intersection in a digital twin, the framework offers actionable insights into intersection efficiency, bottleneck formation, and control optimization. Such system-level understanding is essential not only for infrastructure planning and redesign but also for analyzing CDA-specific applications like cooperative perception, multi-agent tracking, and infrastructure-assisted trajectory planning, making the framework a valuable tool for both research and practical deployment.

\begin{table}[h]
\centering
\scriptsize
\caption{
Agent-level evaluation metrics on V2X and I2I scenarios. ATE = Average Translation Error, ASE = Average Scale Error, AOE = Average Orientation Error, AP@0.5 = Average Precision at IoU = 0.5, AP@Dist = Distance-based AP following the nuScenes evaluation protocol. The best performance for each metric is highlighted in bold.
}
\vspace{-2mm}
\begin{tabular}{c@{\hskip 3pt}|c@{\hskip 4pt}|c@{\hskip 4pt}c@{\hskip 4pt}c@{\hskip 4pt}c@{\hskip 4pt}c}
\toprule
Application & Method      & ATE   & ASE   & AOE   & AP@0.5  & AP@Dist   \\  \hline
\multirow{2}{*}{V2X} 
     & No Fusion   & 0.517 & 0.209 & 0.088 & 0.348 & 0.378  \\
     & Late Fusion & \textbf{0.409} & \textbf{0.190} & \textbf{0.076} & \textbf{0.481} & \textbf{0.507}  \\ \hline
\multirow{2}{*}{I2I} 
     & No Fusion   & 0.430 & 0.221 & 0.153 & 0.487 & 0.515  \\
     & Late Fusion & \textbf{0.262} & \textbf{0.135} & \textbf{0.139} & \textbf{0.592} & \textbf{0.626}  \\ \toprule
\end{tabular}
\label{table:eval_metrics}
\vspace{-3mm}
\end{table}

\begin{table}[h]
\vspace{-2mm}
\centering
\caption{
System-level performance of the CARLA simulator under increasing intersection counts. Percentage changes are shown relative to the previous row.
}
\vspace{-2mm}
\label{table:carla_scalability}
\scriptsize
\resizebox{0.95\linewidth}{!}{
\begin{tabular}{c@{\hskip 6pt}|c@{\hskip 6pt}c@{\hskip 4pt}|c@{\hskip 4pt}c@{\hskip 4pt}|c@{\hskip 4pt}c@{\hskip 4pt}|c@{\hskip 4pt}c}
\toprule
\# Intersections & \multicolumn{2}{c|}{FPS} & \multicolumn{2}{c|}{CPU (\%)} & \multicolumn{2}{c|}{MEM (\%)} & \multicolumn{2}{c}{RSS (MB)} \\ \hline
1 & 1.10 & -- & 288 & -- & 6.7 & -- & 2166 & -- \\
2 & 0.68 & \textcolor{red}{-38.2\%} & 275 & \textcolor{red}{-4.5\%} & 8.4 & \textcolor{green!60!black}{+25.4\%} & 2686 & \textcolor{green!60!black}{+24.0\%} \\
3 & 0.44 & \textcolor{red}{-35.3\%} & 309 & \textcolor{green!60!black}{+12.4\%} & 8.7 & \textcolor{green!60!black}{+3.6\%} & 2803 & \textcolor{green!60!black}{+4.4\%} \\
4 & 0.40 & \textcolor{red}{-9.1\%} & 372 & \textcolor{green!60!black}{+20.4\%} & 9.0 & \textcolor{green!60!black}{+3.4\%} & 2869 & \textcolor{green!60!black}{+2.4\%} \\
\toprule
\end{tabular}
}
\vspace{-5mm}
\end{table}

\noindent \textbf{Agent-Level Performance Analysis.} Leveraging the full-stack CDA capabilities provided by CDA-SimBoost, we design an experiment in which an CDA-controlled ego vehicle navigates through a dynamic intersection environment. During navigation, the ego vehicle interacts with real-world data streams, requiring continuous adaptation of its cooperative perception, planning, and control strategies. In the V2X interaction scenario, we evaluate different perception strategies. Specifically, we assess the perception performance of the ego vehicle under both no-fusion and fusion-based approaches. To demonstrate the infrastructure-centric perception capabilities and advantages of our CDA system, the infrastructure units are equipped with more powerful 64-line LiDARs and cameras. Furthermore, 2–3 IUs are deployed at the same intersection to enable spatially distributed sensing. Table \ref{table:eval_metrics} also presents the perception performance under different strategies in I2I scenarios. The results from this complex interaction scenario validate the framework’s ability to generate meaningful, high-fidelity simulation environments, effectively supporting the testing, training, and refinement of cooperative driving agents across diverse urban settings.

\noindent \textbf{Simulator Scalability Profiling.} 
To evaluate the scalability of the CARLA simulator under varying infrastructure densities, we measure key system-level metrics, including CPU usage, memory consumption, Resident Set Size (RSS) and server-side frame rate (FPS), as the number of intersections increases. As shown in Table~\ref{table:carla_scalability}, the simulator experiences a notable degradation in FPS and a steady rise in CPU and memory utilization as intersection count grows. Overall, the results highlight that while CDA-SimBoost introduces increased computational costs with added intersections, the growth remains within practical bounds. The modest rise in CPU and memory usage—despite the steep FPS decline—demonstrates the simulator's ability to scale up to moderately complex urban scenarios without overwhelming system resources.

\section{Conclusions And Future Work}

% Real2Sim2X marks a significant advancement in bridging real-world traffic environments with high-fidelity infrastructure-driven simulation. By enabling seamless integration of live sensor streams, modular digital twin construction, and dynamic scenario generation within the OpenCDA-InfraX platform, it addresses critical challenges in cooperative autonomous driving-particularly in modeling rare events, managing hetergeneous sensors, and supporting mixed-reality experimentation. Beyond its immediate 

This paper presents CDA-SimBoost, a digital twin and infrastructure-based simulation framework designed for high-fidelity digital twin modeling, challenging synthetic scenario generation, and full-stack CDA verification. The framework integrates real-time and offline data pipelines, a modular digital twin builder, and the infrastructure-centric OpenCDA-InfraX platform to accurately replicate real-world traffic dynamics. It supports scalable development, testing, and benchmarking of CDA algorithms. CDA-SimBoost addresses key challenges in current CDA research—such as real-synthetic data fusion, dynamic scenario generation, and multi-modal data integration—while remaining extensible and adaptable to future infrastructure deployments.

% Looking ahead, future work will focus on expanding CDA-SimBoost to additional deployment locations, with the goal of modeling smart urban city composed of multiple connected intersections. This expansion would enable distributed simulation and testing environments that integrate both virtual and physical assets, supporting hybrid experimentation. Additionally, the platform can be extended to incorporate and validate more advanced CDA algorithms, particularly those involving complex multi-agent coordination and edge-cloud collaborative intelligence. Through continued development, CDA-SimBoost aims to accelerate infrastructure-oriented CDA research and facilitate the deployment of safer, smarter transportation systems.

\bibliographystyle{IEEEtran}
\bibliography{reference_mod}

\end{document}